\documentclass[letterpaper, 10 pt, conference]{ieeeconf}  

\IEEEoverridecommandlockouts                              
\usepackage{graphicx}
\title{\LARGE \bf
Hydra: A Peer to Peer Distributed Training \& Data Collection Framework
}

\author{Vaibhav Mathur$^{*}$ and Karanbir Chahal$^{*}$ 
\thanks{$^{*}$ Equal Contribution. Names are in random order}%
}

\begin{document}

\maketitle
\thispagestyle{empty}
\pagestyle{empty}

\begin{abstract}

The world needs diverse and unbiased data to train deep learning models. Currently data comes from a variety of sources that are unmoderated to a large extent. The outcomes of training neural networks with unverified data yields biased models with various strains of homophobia, sexism and racism. Another trend observed in the world of deep learning is the rise of distributed training. Although cloud companies provide high performance compute for training models in the form of GPU's connected with a low latency network, using these services comes at a high cost. We propose Hydra, a system that seeks to solve both of these problems in a novel manner by proposing a decentralized distributed framework which utilizes the substantial amount of idle compute of everyday electronic devices like smart phones and desktop computers for training and data collection purposes. Hydra couples a specialized distributed training framework on a network of these low powered devices with a reward scheme that incentivizes users to provide high quality data to unleash the compute capability on this training framework. The benefits of such a system is the ability to capture data from a wide variety of diverse sources, something that has been a problem in the current scenario of deep learning. Hydra brings in several new innovations in training on low powered devices, one of which is introducing a fault tolerant version of the All Reduce algorithm. Furthermore, we introduce a reinforcement learning policy to decide the size of training jobs on different machines on a heterogeneous cluster of devices with varying network latencies for Synchronous SGD. The novel thing about such a network is the ability of each machine to shut down and resume training capabilities at any point of time without restarting the overall training. To enable such an asynchronous behaviour, we propose a communication framework inspired by the Bittorrent protocol and the Kademlia DHT. The Hydra network also uses a reward offering similar to the coin offerings of cryptocurrencies to incentivize data collection. It espouses several validation techniques to verify data by providing a novel approach which combines a mixture of neural networks and user feedback.

\end{abstract}

\section{INTRODUCTION}

Deep learning models require a large amount of data to train, they also require this data to be an accurate representation of the phenomenon they're modelling. Creating these datasets is in itself a non trivial task as it requires collection, validation and in some cases annotation, much of which is a manual task presently. Internet scale companies crowdsource these tasks by leveraging their audience's data to create such datasets however annotation and validation is still a bottleneck when the size of data ranges in petabytes. Solutions like Amazon Mechanical Turk \cite{turk} have had some success by anointing human workers to collect and validate data. 

Training models on these huge datasets also takes a substantial amount of time sometimes ranging from days to weeks. Distributed training has found recent successes in bringing down this training time, a major milestone of which was bringing down the training time of the Imagenet database \cite{deng2009imagenet} from weeks to minutes in \cite{jia2018highly}. We propose a decentralized peer to peer framework coined Hydra for machine learning dataset creation and distributed training. It makes these two functionalities interdependent to incentivize data collection thereby ensuring a steady influx of high quality data for training purposes. Hydra trains models on a heterogeneous cluster of varying device types and network conditions unlike a data center where devices are generally homogeneous and are connected with a high bandwidth and low latency network. An added challenge in training on Hydra's cluster is the unpredictable nature of Hydra nodes due to their unreliability as they can shut down and come back online arbitrarily. 

Training on a cluster of low powered devices with high bandwidth low latency connections has been recently explored by a paradigm known as Federated Learning (FL) \cite{federated}. A core feature of FL is preserving the privacy of user data while training by adopting a zero user data transfer policy and encrypting weight sharing using a variety of techniques on a master slave setup \cite{ryffel2018generic}. Conversely, Hydra proposes a peer to peer paradigm where weight exchange uses the all reduce procedure thus substantially reducing the role of the coordinator. Hydra trains on data considered public hence removing the need to encrypt it. This allows for a data distribution policy that seeks to maximize training efficiency according to device type and network connections unlike FL.

We divide this paper into two major sections.  The first section delves into the core peer to peer structure of Hydra by mapping out it's individual components, supported functionalities and explains how data uploading is incentivized. The second section looks at the distributed training framework within which individual sections explore several optimizations and novel job placement techniques to  bring distributed training to low powered devices. Finally we end the paper with the conclusion which discusses the implications along with the challenges with Hydra and summarizes the discussion.

\section*{Core Structure}

Hydra is a peer to peer network of nodes that offers the functionality to build crowd sourced datasets and train machine learning models. The Hydra network has two major components, the peer node and the bootstrap server.


\begin{itemize}
  \item \textbf{Peer Node}: The peer node is a fundamental component of the Hydra network. Every user who joins the Hydra network is a peer. Peers connect to one another and provide the functionality of creating new datasets, contributing to existing datasets and training machine learning models in a distributed fashion. All peers are homogeneous and non-Byzantine. Every peer on the network is given a unique $peer\_id$. The $peer\_id$ acts as the identification token of that peer in the network.
  \item \textbf{Bootstrap Servers}: These are a group of centrally maintained servers which are tasked with the responsibility of inducting new peers into the Hydra network. They provide the $peer\_id$ for a new peer and trigger a peer initiation routine to inform other peers of the existence of the new peer. Bootstrap servers also keep a track of the location of the trackers of datasets that are currently hosted by Hydra. This information is important for peers to access or start contributing to an existing dataset and is further explained in the coming sections. It is imperative to note that bootstrap servers are always available unlike peers and do not participate in the core functionality offered by Hydra, they are simply a gateway for prospective peers to join the network and provide information on datasets that Hydra currently offers.
\end{itemize}

\section{Distributed Hash Table}


For a peer to communicate with and locate different peers on the network, it needs to store their physical address. Since the number of peers can be extremely large, it is not feasible for a single machine to store the information of every other peer in it's memory. For this purpose, Hydra uses a variant of the Kademlia Distributed Hash Table (DHT) \cite{kademlia} to store the location information of different peers across the network. We break down the explanation of the functioning of the Hydra DHT into three parts. First, we explain the general structure of the DHT and it's underlying lookup tables. Second, we explain the implementation details behind various operations offered by the lookup table and lastly we delve into how the DHT is used for various Hydra operations such as the \textit{Find Nodes} and \textit{Request  Data} in the next section.


\subsection{General Structure}

The Hydra DHT is composed of a number of lookup tables, each peer has a lookup table ($DHT_{peer\_id}$) stored in it's memory. The $DHT_{peer\_id}$ of a peer prioritizes storing the location information of peers closest to itself, this degree of closeness is defined by  the XOR distance between two $peer\_ids$.
\begin{equation}
D(P_{1},P_{2}) = peer\_id_{P_{1}} \oplus peer\_id_{P_{2}}
\end{equation}

The number of bits in the $peer\_id$ is denoted by $N$, we use $N=256$ in our implementation. Each $DHT_{peer\_id}$ has N keys, the value for each key is a list of maximum size $M$. These lists store the location information of various peers, a single entry of a peer in the $M$ sized list contains the physical address and the $peer\_id$ of the stored peer.

\subsection{Operations}

The $DHT_{peer\_id}$ lookup table supports two operations- insertion and lookup.
\begin{itemize}
\item \textbf{Peer Insertion}:
To store the information of a peer $P_{1}$ with $peer\_id_{P_{1}}$ in the lookup table of peer $P$ with $peer\_id_{P}$, $D(P,P_{1})$ is calculated. Next, the position of the first non zero most significant bit ($i$) of $D(P,P_{1})$ is used as a key to query the lookup table to get $List_i$. If $len(List_i)$ \textless $M$, then the peer information can simply be appended to the list. If the list has a size equal to $len(List_i)=M$, peers stored in that list are checked for livliness through heartbeat messages. If some peers are offline, the information for peer $P_1$ replaces any one of the offline peers in the list. If none of the peers are offline, the insertion request is rejected. This is done as Hydra will always prefer to exploit old reliable peers over exploring new ones.

\item \textbf{Peer Lookup}:
This operation is used to retrieve the location information for some $peer\_id_{P_{1}}$ from the lookup table, $DHT_{P}$. This operation is similar to the Peer Insertion operation, the only difference being that after the $List_i$ is obtained, the $peer\_id_{P_{1}}$ is searched for in the list. If the  $peer\_id_{P_{1}}$ is present in $List_i$, the operation fetches the physical address of $P_{1}$ from the $DHT_{P}$. If the $peer\_id_{P_{1}}$ is not found, a null response is returned.

\end{itemize}

\textbf{Intuition}: The intuition behind using this methodology of storing values based on the first non zero most significant bit (MSB) of the XOR of two $peer\_ids$ is important to understand. Using the first non zero MSB indicates the degree of closeness of the peer requesting an insertion with the host peer node. For example, if a peer $A$ has $i=4$ for peer $B$, that entails that the peer $A$ is at least as close as $N/16$ other nodes on the network to peer $B$. Similarly, if peer $C$ has $i=2$ with $A$, it is as close as $N/4$ other nodes on the network, hence making peer $A$ closer to peer $B$ than peer $C$. Every time we move a bit towards the left, we are effectively discarding half of the remaining number of nodes on the network. This property enables Hydra to find any peer on the network given it's $peer\_id$ in $O(log(N))$ time where $N$ is the total number of peers on the network. A higher value of $i$ denotes a higher probability of the peers in $List_{i}$ knowing about the requested peer. 

If one was to think of all peers of Hydra represented in a graph, a peer's $peer\_id$ would define the location of that peer within the graph hence peers with similar $peer\_ids$ will cluster closer to each other.

\section{Hydra Operations}

\subsection{Find Node}

The \textit{Find Node} operation is used to discover the location information of some peer given it's $peer\_id$. This is a fundamental operation of Hydra and has various usages such as finding a peer to appoint for some training job or request for some chunk of data among other uses.

Let's assume that $P$ with $peer\_id_{P}$ wants to communicate with $P_1$ having $peer\_id_{P_1}$. Initially, $P$ performs the \textit{Peer Lookup} functionality on it's $DHT_P$. If the operation is successful, the address is retrieved and used by $P$ to establish a connection with $P_1$.
On failure of the lookup, $P$ generates a $k^{th}$ sized list ($L$) of peers closest to $P_1$. This degree of closeness is calculated using the same formula as (1). After computing $L$, a \textit{Find Peer} request is sent to each peer in $L$, these request query the peers for the address of $P_1$. If found, the peer returns the location information of $P_1$ and if not, it returns a list of the $k$ closest peers. On receiving $k$ number of $k$ sized lists, peer P builds a refreshed $L$ of peers closest to $P_1$. This cycle is continued until $P_1$ is found or when the peers in $L$ are not closer to $P_1$ than the peers in the previous list. 

It is also important to note that every time a peer performs a \textit{Peer Lookup}, it also performs a \textit{Peer Insertion} of the requesting peer. This is done to augment the peer's knowledge of the Hydra network. In other words, peers gets smarter every time a \textit{Peer Lookup} is called. The \textit{Peer Insertion} call is performed asynchronously so that it doesn't affect the execution time of the \textit{Find Peer} operation.

This operation is computed efficiently due to the way the Hydra DHT is structured. The \textit{Find Peer} operation runs in $O(logN)$ time where $N$ is the total number of peers. The method Hydra uses to get the $peer\_id$ of the peer to be located will be explained in later sections that look into the creation and contribution to datasets.
\subsection{Induction of New Nodes}
 To join the Hydra Network, a prospective peer is granted a $peer\_id$ by a bootstrap server. After receiving it, the prospective peer fires a \textit{Find Node} request for it's own $peer\_id$ to the bootstrap server. The intuition behind this operation is to make the peers of Hydra aware of the entry of a new peer into the network by adding it's $peer\_id$ in their hashtables. The operation also has the added benefit of informing the new peer about the location information of various peers in the network and help it populate it's $DHT$.
 
 \subsection{Creating a New Dataset}
 Creating a new dataset involves a number of steps. First, an SHA256 hash of the title of the dataset is created which is denoted by $H$. Second, a \textit{Find Node} query is fired for $H$ to discover the closest peer to the value of $H$, this peer will track all changes and meta data information about the dataset. We shall call it a tracker as it is similar to a tracker found in bit torrent systems \cite{bittorrent}. Finally, the dataset title is sent to the bootstrap server to append to it's list of existing datasets in Hydra. The bootstrap server distributes this data to all other bootstrap servers to replicate this information. Once this process is complete, one can start contributing data to a dataset. Hydra replicates the tracker and ensures it's fault tolerance as it is an important component of the dataset, trackers are explained extensively in the next section
 
 \subsection{Contributing to an Existing Dataset}
 
 To contribute to an existing dataset, first the peer should have a list of existing datasets to choose from. These lists are maintained by bootstrap servers. Once this list is received, a peer can choose the dataset it wants to contribute to. Once a peer has chosen a dataset, it's SHA256 code which is denoted by $H$ is used to find the tracker for that dataset with the \textit{Find Node} request. Once the location information for a tracker is found, the peer informs the tracker of the information it wants to contribute. More specifically, it sends the size of the data it is contributing and meta data such as filenames along with the peer's location information. This information is stored on the tracker against the dataset key. Hence, the tracker keeps a key value store where the key is the dataset title and the value corresponds to a list of metadata which informs about the filenames, filesizes and the peer(s) that have contributed and downloaded the dataset.
 
 \subsection{Downloading from an Existing Dataset}
 
 For a peer $P$ to access the data of a particular dataset, it requests the dataset tracker for the list of peers that host the actual data, $L^{peers}$. This list contains the peers that have contributed or downloaded the dataset. After receiving this list, $P$ parses it and fires a request for the dataset's data to all the peers in $L^{peers}$. This operation starts the file transfer of the actual dataset data from the source peer to requesting peer. After downloading this data, $P$ effectively has a copy of some portion of the dataset. Hence, after it initiates the download of the dataset, $P$ requests the tracker to add it to $L^{peers}$, thereby informing other peers of an alternative source they can download from. Similar to a torrent like system the replication of a file is commensurate with the number of peers downloading it, this information can later be used for analyzing various user and dataset trends in Hydra.
 
 Once a downloading peer puts itself up for seeding, it is rewarded commensurate to the amount of bytes transferred by it to another peer. This is further explained in the \textit{Incentivizing Users} section. Note that the tracker simply keeps a track of original contributors and downloading peers to minimize the amount of information transferred between the tracker thereby reducing  it's load. Other decision like downloading strategy and rewards allocation are handled by a "tit for tat" strategy between two peers with no governing entity.
 
 There are several algorithms espoused by the bit torrent protocol for downloading files which intelligently decide where and how to download a file from after parsing the $L^{peers}$, however this is considered as out of scope for this work.
  In this paper, we assume the Hydra network is secure with no malicious peers, although it is a possibility, this is also considered out of scope for this work. Delving into download, reward and security strategies are planned to be addressed in future work.
 
 \subsection{Triggering a Training Job}
 
 For a peer ($P$) to fire a training job on Hydra, it first retrieves a list of peer nodes that are available for training. This list is prepared using a variety of metrics such as the limit of compute that $P$ is offered which is dependant on how much Hydra coin (described in later sections) $P$ wants to spend. This list is also computed using factors such as degree of closeness, compute capacity and network conditions connecting these peers. Once a final list of devices is prepared, $P$ sends a request to the distributed training framework to start the training where $P$ acts as an orchestrater for the training and tracks it's progress. It also receives the final set of weights of the trained model. The exact operation of how the distributed training framework operates is explained in later sections.
 
\section{Multi Trackers}

Trackers are an interesting concept in Hydra. One can think of them as representatives of a dataset. As datasets are a crucial commodity in Hydra, trackers need to be properly replicated such that there is minimal loss of information of a dataset when a tracker goes down. We propose using a novel \textit{Multi Tracker} approach to replicate trackers by leveraging the Raft \cite{raft} protocol (described later). Trackers are replicated across a cluster of  nodes and state is maintained in a consistent, available and fault tolerant manner through Raft. Each cluster of nodes has a leader through which all client communication takes place. We detail how this is achieved in detail by exploring the subroutines that are run after a tracker gets initialized:

\begin{itemize}
    \item \textbf{Maintaining a Fixed Number of Replicas}: A multi tracker approach espouses the use of a group of replicated trackers that maintain a consistent state among them. A leader of a tracker group is a node through which the client interacts with and all state changes pass through from. This leader is also responsible for maintaining a fixed number of replicas, say $N$. Hence, if a replica goes down, the leader has to choose nodes to replace it. The list of nodes to anoint as replicas are computed by using the \textit{Find Nodes} request for the SHA256 hash of the dataset. The top $N-1$ nodes out of this list except for the leader are anointed replicas for that dataset. 
    
    \item \textbf{Maintaining State between Replicas}: State management and consensus is performed using the Raft protocol which is described in the Distributed Training section of this paper. The leader of the cluster of nodes handles all client requests and performs the state changes on at least a majority of replicas before returning a confirmation response to the client. This ensures that the data can be preserved in case of a crash. Whenever a replica goes down, a prospective replica node is identified and brought up to date with the dataset changes before incorporating it into the group. If a leader goes down, the Raft's leadership election procedure anoints a new leader whose first priority is to get the number of replicas back up to $N$.
    
    \item \textbf{Informing the Network of Leadership Change}: Everytime a leader is changed, the new leader needs to inform the peers of Hydra of the new location of a dataset. Simply computing a SHA256 hash of the dataset to find the tracker can lead to wrong results as Hydra has various peers joining and dropping off unpredictably resulting in a dynamic graoh structure due to the constant $peer\_ids$ generation. We use bootstrap servers to keep a track of the active leaders of a dataset and this information is accessed by peers for contributing to or downloading a dataset.
    
    \item \textbf{Rebooting tracker state}: A dataset's information can be completely lost if all the trackers for that dataset are down. This is a plausible situation in Hydra due to the unreliable nature of it's nodes. To partially mitigate this problem, the peer that creates the dataset periodically takes a snapshot of the metadata of a dataset from the leader tracker and stores it on it's memory. It regularly contacts the bootstrap servers for a livliness check of it's dataset's trackers. If none of the trackers are online, this peer elects a new tracker after copying over the dataset's metadata. Once this new tracker is setup, it follows it's regular operation of electing and maintaining replicas. There is a possibility that the metadata contributed can be old, this is partially mitigated by comparing the data of other downloaders of that dataset when they try to access the tracker. The tracker updates it's metadata accordingly seeking to obtain the latest version of that dataset in the event of a multi tracker crash. 
\end{itemize}

\section{Incentivizing Users}

For a system like Hydra to work effectively, it needs to build a framework for a virtuous cycle of injection of data and training on that data. We seek to tightly couple the two by unlocking training capability proportional to the amount of \textit{high quality} user contributions towards building and maintaining a high dataset. To achieve this, we propose Hydra coin, inspired by the concept of cryptocurrency \cite{nakamoto2008bitcoin}. The coin determines the amount of training compute a user is allowed on the network. There are a myriad of ways to earn this coin, these ways seek to ensure that Hydra injects high quality verified data into it's datasets.

\begin{itemize}
    \item \textbf{Data Contribution}: Hydra coin is rewarded to the user proportional to the amount of data they contribute. An exact coin exchange rate for the size of data can be decided by the implementer. There is also a mechanism that subtracts coin from a user's treasure chest if the data they upload is irrelevant, duplicate or malicious. Users are also rewarded extra coin if they contribute valid data to a variety of datasets to encourage diversity of data.
    \item \textbf{Data Validation}: Hydra espouses a manual data validity check that is similar to a crowd sourced Amazon Mechanical Turk \cite{turk}. Users are rewarded Hydra coin if they validate data. An example of that could be validating that all images in a cat dataset have cats in them. Users are rewarded coin proportional to the amount of data they validate, if some malicious or irrelevant data is found we penalize the original contributor to discourage such behaviour. In the future, Hydra could use some form of an anomaly detection algorithm based on deep neural networks that can raise concerns over some data items, similar to a spam detector. Identifying duplicate data could also be possibly automated in future work.
    \item \textbf{Data Annotation}:
    Some datasets like those for object detection, segmentation and image classification require annotation. Users are rewarded coin proportional to the amount of data they annotate with useful targets. These targets are necessary for various tasks such as classification, regression etc. A dataset that supports a greater number of tasks is more versatile than a dataset which doesn't. Data annotation helps increase the quality of a dataset and gives it additional features that originally weren't there. 
    \item \textbf{Training}: Hydra coin is awarded proportional to the amount of compute the user contributes to the network. We propose a Virtual Compute Unit (VCU) that acts as the standardized unit of compute. To calculate the amount of VCU for machine $m$, we use the following equation:
    
    \begin{equation}
        VCU_{m} = sigmoid(t_{b}-t_{m})*A 
    \end{equation}
    
    where $t_b$ is time taken to run the model for a single data point on the bootstrap server, $t_m$ is time taken to run the model for a single data point on the machine $m$ and $A$ is the amount of the data $m$ trains in one step. For instance, the VCU for a bootstrap server would be $0.5$. Hydra coin is awarded when a machine trains a single batch and successfully communicates it's weights.
\end{itemize}

Users can expend their Hydra coin to train their neural networks on the distributed training framework. The exact value of Hydra coin in terms of VCUs is left up to the reader, we recommend using a configuration which maintains a healthy balance between both training and data collection. We maintain a direct correlation between the value of a coin and the VCU. A more powerful heuristic for Hydra coin would be a metric that understands the current capacity of the network and varies value, this is however considered as future work.
\section*{Distributed Training Framework}

Distributed training frameworks have been usually constructed keeping a high performance cluster in mind. Hydra differs from them as it brings distributed training to a resource constrained scenario. Generally, distributed training frameworks consist of an optimization algorithm and a data transfer specification. Among optimization algorithms, Stochastic Gradient Descent (SGD) with Nestrov's Momentum \cite{ruder2016overview} is a popular choice and is quoted in used in most distributed training research. SGD is further broken down into Asynchronous  SGD \cite{dean2012large}  and  Synchronous SGD \cite{goyal2017accurate}. Asynchronous SGD uses a lazy gradient upgrade policy where after a node completes it's computation of gradients, it sends them to the master server and continues on to the next batch of data. Whenever the master gets a new update of gradients, it broadcasts them over the network informing the nodes about the new gradients. These nodes apply the updated gradients to their set of weights before performing the training on the next batch of data. This decoupled nature leads to numerous problems in Async SGD, the major ones being divergence during training and failure to reach the test accuracy benchmark of a model trained on a single machine. Research has eschewed Async SGD and generally accepted Synchronous SGD as the state of the art optimization algorithm. Inspite of the synchronization barrier, Sync SGD has proven to provide a reliable corner stone to several state of the art training frameworks \cite{sergeev2018horovod,jia2018highly}. 

An important component of a distributed training network is the collective communication primitives used to transfer gradients around the cluster. The all reduce algorithm from the world of High Performance Computing (HPC) has found great success in training frameworks \cite{gibiansky2017bringing} as it allows for the removal of a master server and optimizes bandwidth utilization. In addition to the all reduce, recent efforts to speed up training have tried increasing batch size, training has been made possible for batch sizes upto a size 64,000 \cite{jia2018highly, ginsburg2018large} by using various modifications. It is observed that increasing batch size naively, leads to divergence in training or worse validation accuracy. Techniques such as linear scaling of learning rate \cite{goyal2017accurate} and layer adaptive learning rates \cite{ginsburg2018large} have unlocked scaling of batch sizes leading to training times for the an image classification model on the Imagenet dataset to be as low as 4 minutes \cite{jia2018highly} without any loss in validation accuracy. It has also been empirically proven that increasing the batch size is equivalent to decaying the learning rate \cite{smith2017don}. Such advances have bought distributed training to a moderately mature state. 

Hydra ports several ideas from the above research, however it introduces a novel approach to training in a resource constrained environment. More specifically the following five modifications are proposed
\begin{itemize}
    \item  A fault tolerant all reduce communication collective.
    \item A modified SGD that supports training in the ever changing environment of the Hydra network that supports nodes connecting and disconnecting frequently.
    \item A gradient compression methodology for transference of gradients in low bandwidth networks.
    \item A reinforcement learning policy which decides the placement of parts of the overall batch size on the heterogeneous cluster of devices.
    \item A mixed precision training approach that supports extremely large batch sizes to maximize scalability, arithmetic throughput, lower memory utilization and optimize network bandwidth.
\end{itemize}

\section{Peer to Peer Synchronous SGD}

Synchronous Stochastic Gradient Descent is a popular algorithm of choice to use in distributed training. It has been used as a backbone to build fast distributed systems that have lowered the state of the art considerably over the years \cite{sergeev2018horovod, jia2018highly}. Synchronous SGD guarantees the replication of the validation accuracy found on a model trained on a single machine provided that the batch size is scaled using techniques like the linear learning rate rule \cite{goyal2017accurate} or LARS \cite{ginsburg2018large}. Asynchronous SGD is another alternative but presently it has numerous problems like stale gradients, inferior convergence accuracy and in some cases divergence. It has been observed that due to these fallacies the entire training procedure needs to be restarted in some cases \cite{chen2016revisiting}. Asynchronous SGD requires a reliable environment where machines complete their computation and share gradients  at around the same time. This requirement does not bode well with the Hydra setup where nodes are extremely unreliable. Hence, the Hydra system espouses the use of Synchronous SGD for training deep learning models.

Synchronous SGD is given as follows
\begin{equation}
 w_{t+1} = w_{t} - \eta \frac{1}{n}\sum_{x\in B} \nabla l(x, w_t)
\end{equation}

where $w_{t+1}$ are the weights computed for the current batch, $n$ is the number of training examples in the mini batch and $\nabla l(x, w_t)$ are the gradients computed for a particular training example.

Synchronous SGD is in theory quite similar to the SGD procedure on a single machine, different chunks of data are allotted to different machines which run the forward and backward pass to compute the gradients for those individual chunks. After gradients are computed, machines all across the cluster communicate with each other to get the new updated set of weights. Each machine shares its gradients with the master server, this server averages all the gradients received and sends the new gradients to each machine as updated weights. As common in modern distributed training frameworks, Hydra discards the master slave pattern for the all reduce algorithm. The all reduce has the benefit of being friendly to a peer to peer setting and also has been observed to provide better utilization of network bandwidth. 

After machines receive their new weights, they can start processing the next chunk of data. One can think of a concatenation of all these chunks across the machines as one  \textit{large} mini batch. The reason Synchronous SGD performs well is by leveraging extremely big mini batch sizes which results in lower overall parameter updates. Batch sizes have ranged from 8,000 to even 64,000 enabling training datasets like Imagenet in as low a time as 4 minutes \cite{jia2018highly}. The hardware setup for these results have powerful GPUs ranging from 256-2048 in number. 
Hydra has a lot of nodes that can shut down and come online at arbitrary times. To enable training on such a system we propose certain addition to a vanilla training framework to help port Synchronous SGD to the peer to peer environment of Hydra. 
\begin{itemize}
    \item To trigger a training job, Hydra uses a selection algorithm (described in later sections) that selects a list of machines out of an available machine pool. These evaluation of machines to be selected takes certain parameters into account that range from ping time, compute availability and degree of closeness to each other. The last point is particularly salient as Sync SGD offers optimal performance in a setting where peers are physically closer to each other as that provides minimal communication latency to balance the computation and communication trade off. After the list of trainable nodes have been decided upon, the training can commence.

    \item Keeping backups of nodes is essential to recover from machine failure. We use a replica based cluster that ensure fault tolerance during training (next section).
    
    \item In case of multiple machine failure where both the machine and it's replicas fail, the training continues on by saving that failure information on the node that triggered the training. This machine keeps track of the chunks of peers that have or haven't provided their gradients. If for some reason a chunk of data could not be computed in the current mini batch, it is sent as part of the next mini batch. The machine informs the rest of the training nodes of the removal of a cluster which allows them to complete their All Reduce procedure. Hence, even through failure the training system shall make parameter updates and work towards reducing the loss. It is important to note that the machine which triggers a training job has to stay online during the duration of the training. Better ways of maintaining state are planned to be addressed in future work.
 
\end{itemize}

Hence, through these modifications we provide a peer to peer compatible version of the Synchronous SGD algorithm that supports nodes arbitrarily dropping on and off. The next sections look into how this fault tolerance is achieved in greater detail.

\section{Fault Tolerant All Reduce}

The function of the All Reduce is to aggregate and reduce the data present on each machine and then store the final result on each machine in the cluster. All Reduce functions without a master server pattern and instead espouses a peer to peer networking paradigm. The benefits of this are better bandwidth utilization and lower latency. The All Reduce was first introduced by \cite{thakur2005optimization} for the task of gradient accumulation required in Synchronous SGD. It is a collection collective primitive used in high performance computing landscapes.

We propose a modification to the recursive halfing and doubling algorithm present in \cite{thakur2005optimization} to make it resilient to machine failures. As originally constructed, the All Reduce procedure needs to be restarted each time a machine in the cluster crashes or shuts down. We select the the recursive halfing and doubling all reduce algorithm to build fault tolerance on top of as it has been found to provide speed gains of nearly 3x over other ring all reduce counterparts \cite{goyal2017accurate}. The speed up is due to the tree based communication strategy that allows data to be reduced and stored in $logN$ steps instead of $N$ steps, where $N$ is the total number of machines in the cluster. However, our modification is independent of the type of all reduce procedure and can work with other variants as well.

Machine failure is a common problem in distributed system, the obvious choice to recover from machine failure is to keep replicas. Maintaining state among a group of replicas in a fast and secure way is however a non trivial problem. There have been several algorithms proposed to address state maintenance in replicas, some of the popular ones being Paxos \cite{paxos} and Raft \cite{raft}. We propose combining the properties of Raft with the All Reduce to build fault tolerance. Raft \cite{raft} is a Multi Paxos \cite{paxos} based distributed consensus system that has been proved in production environments dealing with global scale.
We propose the following modification in the All Reduce algorithm. Among a set of $M$ nodes, $N$ groups are formed with $M/N$ machines each. The all reduce procedure is carried over these $N$ groups. The Raft protocol operates over $M/N$ machines on each of these $N$ groups.

\section*{Raft}

Each node in a Raft \cite{raft} cluster has three states

\begin{itemize}
    \item \textbf{Follower}: This node follows the state changes of the leader, whenever data is changed on the leader, the leaders sends the changes over to the followers. Only after a majority of followers have performed these changes, does the leader confirm the state change for that particular cluster. Data is exchanged on top of heartbeat messages that the followers and leader exchange after a fixed time period to check livliness.
    \item \textbf{Leader}: The machine that tracks all the state changes and receives the data update requests is known as the leader. It interacts directly with the client and is elected by the Raft protocol through a process called leader election.
    \item \textbf{Candidate}: There might come a time when the leader of the group is down. Candidate nodes are follower nodes that announce their candidacy for being the next leader. Follower nodes become candidate nodes if they haven't received a heartbeat message from the leader in some time, this time interval is randomized between 150-300ms. Candidate nodes ask for votes and whichever candidate receives the majority of votes soonest, becomes the leader. A candidate if asked for a vote will vote for the requestee node if it hasn't voted for some node in this election.
\end{itemize}

\subsubsection{State Changes}

Once some change to the data is requested for a node, the leader gets the request. The leader performs the data modification and sends the same request to all the replicas. Only after a majority of requests are confirmed by the replicas, does the leader commit the change and send back a confirmation to the client.

\subsubsection{Leader Election}

There might come a time, when the leader is down. In such a scenario a new leader must be elected. Raft uses a randomized timeout after which nodes submit their candidacy for leadership. After being available for candidacy, the candidate request votes in the form of vote messages to each replica. 
If a node receives a vote request and hasn't voted before in this election term, it votes for the requestee node. Once a majority of nodes have voted for a leader, a new term is begun with the new leader and requests are served.

\subsubsection{Partition Tolerance Recovery}

If the cluster gets divided into two parts, it could have 2 leaders at a single point of time. Once that partition is healed, Raft dictates that the leader which has the most nodes listening in is allotted the position of the new leader, the previous nodes have to sync up with the new leader, after which they are allowed to join the network.

\subsubsection{Recovery from Split Vote}

Sometimes a split vote can occur during leader election. Imagine a scenario when two nodes timeout at the same time and announce their candidacy, if the votes they receive are equal the cluster waits for the next election timeout to solve this split vote. Raft uses a randomized timeout for this very reason, split votes are resolved quickly if such a scenario arises.

\section*{Recursive Halfing Doubling All Reduce}

The recursive halfing and vector doubling algorithm \cite{thakur2005optimization} works for $O(logN)$ steps.
It is a combination of two subroutines, the vector halfing scatter reduce and the vector doubling all gather. The scatter reduce after completion stores a part of the final result on each machine and the all gather is responsible for accumulating this data on each machine. Both of these algorithms have a time complexity of $O(log(N))$ steps, the scatter reduce algorithm is explained first below.

\subsection{Scatter Reduce}

At each step data is exchanged between machine $i$ and $i + B$. The value of $B$ initially is $P/2$ and after each step is divided by 2, where $P$ is the total number of machine participating in the Raft cluster. The exchange of data is performed by halfing the data vector into 2 parts- one for sending data and one for receiving data. The two pairs of machines divide this data in inverse ways. In other words, if the top half of data of machine A is meant for sending and bottom half for receiving, machine B will use the opposite configuration. After the data exchange is completed, the data received is reduced with the original data at that position. Also the new vector to be used from now on will be the reduced data, effectively half of the vector operated on before, hence validating the term "vector halfing" in the algorithm name. This process continues on for $O(log(N))$ , after which each machine has some part of the final reduced vector. The all gather phase is next where all of this reduced data is accumulated on each machine.

\subsection{All Gather}

The all gather works by reversing the steps followed in the scatter gather. It doubles the vector and works by communicating between machines in distance $B$, with $B=1$ and is doubled after each step for $O(log(N))$ steps. At each step, data is exchanged and the final vector is doubled in size by concatenating the data being received with the reduced data present on the machine.

\section*{Combining Raft and All Reduce}

When the scatter reduce is taking place, if a machine fails or takes a long time to respond, the whole process gets held up. In most cases, the entire procedure has to be restarted. We propose the following modifications

\begin{itemize}
    
    \item Each participating node in the All Reduce procedure will have a set of replicas listening in.
    
    \item These replicas will elect a leader which listens in for state changes during the All Reduce procedure. Once a change is requested, the replicas perform the state change and respond with the intended result. The leader relays this back to the client. The client keeps polling the leader with heartbeat messages to check livliness. 
    
    \item In the case of machine failure a new leader will be elected and the request will be sent again by the client after it is informed of a leadership change.
    
    \item It is non trivial to find the identity of active leaders of each node cluster as leaders might go down unpredictably. To propagate the identity of the newest leader of a node cluster, we use a central server (the initiating node of training job) that keeps track of the active leaders of each cluster. Therefore before any communication for the all reduce occurs, the latest leader info is retrieved via this central server adding a small overhead.
    
\end{itemize}

This setup has a number of advantages. Firstly, the all reduce procedure will be unaffected in case of replica failure. In the case of leader node failure, the operation will simply be needed to be repeated again after a new leader is elected instead of restarting the whole procedure. Recent work verifies our strategy of maintaining replicas as backup workers which allows for a faster overall training time \cite{chen2016revisiting}. This kind of setup works for a system like Hydra which has numerous unreliable nodes.

\section{Hydra Distribution Policy}

The challenge in training on a decentralized network is the heterogeneous nature of devices which are interconnected with varying network latencies. This environment is unlike data centers such as the Google Cloud Platform and Amazon Web Services where individual devices are homogeneous and the most part reliable and connected with a low latency and high bandwidth network. Hydra uses machines of varying compute capabilities and network conditions to train machine learning models. Training using Synchronous SGD on this cluster with each machine using a constant batch size can lead to under utilization of machines with a greater compute capacity. Conversely, machines with a smaller compute capability might be overwhelmed by the training job.

To solve this problem, Hydra proposes a novel approach. It proposes allocating different batch sizes that are proportional to the compute and network capacity of this cluster of heterogeneous devices. There are several problems that need to be solved to achieve such a placement configuration.

\begin{itemize}
    \item Devices are connecting and disconnecting constantly, a placement policy will need to keep track of the constantly changing compute power and network structure of the cluster.
    
    \item The placement policy should be capable of implicitly or explicitly modelling the optimal data distribution strategy for training on each device i.e it must calculate the optimal batch size for a device that is commensurate with it's compute and network capabilities.
\end{itemize}

We propose using a Reinforcement Learning (RL) solution to figure out the optimal configuration to attain the least training time. An RL solution will determine an optimal policy to distribute a dynamic batch size to each device. The reward function is a function of the overall training time for a single batch size over the network, the higher this time, the lesser the reward and vice versa. Due to the algorithm's continual nature, it will get better with time and may even discover optimal unconventional positioning strategies that wouldn't be otherwise explored by a deterministic algorithm. There is a history of RL algorithms finding optimal unconventional solutions in game playing \cite{silver2017mastering}, automated neural architecture search \cite{automl} and model compression \cite{rlcompression} which is why this approach made sense. The use of RL for distributed training has been explored for the case of network node placement on device \cite{rltensorflow} (model placement), however our work chooses to automate the distribution of batch sizes in Synchronous SGD and tackles a different aspect of distributed training (data placement).

The RL policy uses two parameters to output a placement configuration- latency time during communication between machines and the compute power for each machine. The compute strength is calculated by running a test model on each device and calculating the time it takes for the model train a single batch of data, this is termed as $C$ for a device. The GPU or RAM size of each device is also sent to the coordinating node to determine the optimal size of mini batch that device can compute, this size is given by $g$.

The network needs a matrix $M$ of size $k*k$ as input. Each value $M_{ij}$ corresponds to the latency time it takes for device $i$ to communicate with device $j$. A vector $V$ is also constructed of size $k$, where $V_k$ denotes the $C$ of the $k^{th}$ device. Another vector $S$ is constructed of size $k$, where $S_k$ denotes $g$ or memory size of the $k^{th}$ device.
Matrix $M$, vector $V$ and vector $S$ are concatenated and fed into a Convolutional Neural Network (CNN). An output is a vector $D$ of size $k$ where $D_k$ represents the mini batch size of machine $k$. The loss computed is compute through a policy gradient algorithm which uses the latency time, $L_t$ of the configuration mentioned in $D$, as the reward function of the algorithm. The vector that the controller predicts can be viewed as a list of actions $a$ to allot the various loads to each job. More concretely, to find the optimal architecture, we ask our controller to maximize its expected reward, represented by:

\begin{equation}
J(\theta_{c}) =
E_{P(a ;\theta_{c})}[L_t]
\end{equation}

Since the reward signal $L_t$ is non-differentiable, we use a policy gradient method to iteratively update $\theta_c$. In this work, we use the REINFORCE rule \cite{Williams1992}:

\begin{equation}
\nabla_{\theta_c} J(\theta_{c}) = E_{P(a ;\theta_{c})}[ \nabla_{\theta_c}log(P(a;\theta_{c})L_t)]
\end{equation}

The above update is an unbiased estimate for our gradient, but has a very high variance. In order to reduce the variance of this estimate we employ a baseline function:

\begin{equation}
    \nabla_{\theta_c} J(\theta_{c}) = E_{P(a ;\theta_{c})}[ \nabla_{\theta_c}log(P(a;\theta_{c})(L_t-b))]
\end{equation}

As long as the baseline function b does not depend on the on the current action, then this is still an unbiased gradient estimate. In this work, our baseline b is an exponential moving average of the previous job schedules.

\section{Training Framework}

Hydra uses several techniques to train networks in an efficient and parallel manner. Training with Synchronous SGD has largely been accelerated by using large batch sizes \cite{goyal2017accurate}\cite{ginsburg2018large}\cite{jia2018highly}. Training with large batch sizes leads to the model requiring lesser parameter updates to converge  which in turn leads to a dramatic speed up in training. It has also been empirically observed that increasing batch size is equivalent to decaying learning rate \cite{smith2017don} thus providing a viable alternative.

Unfortunately, naively increasing the batch size results in divergence issues and a "generalization gap" \cite{jin2016scale}. A model trained on a large batch size has lower validation accuracies on the test set when compared to a model trained with smaller batch sizes. There have been a number of methods that have been proposed to scaling batch sizes such as \cite{goyal2017accurate}, \cite{ginsburg2018large} and \cite{jia2018highly}. Hydra uses the LARS algorithm \cite{ginsburg2018large} which espouses the use of a learning rate for each individual layer of the network. Layer wise adaptive learning rates were introduced from the observation that different layers of the network learn at different rates. The local learning rate of a layer is computed by taking the ratio between the norm of the gradients and weights of the layer and weights it with a hyperparameter and the global learning rate. The local LR  $\lambda^l$ for each layer $l$ is computed as follows:

\begin{equation}
  \triangle w^l_t  =  \gamma* \lambda^l * \nabla L(w^l_t) 
 \end{equation}
 where $\gamma$ is a global LR.  Local LR  $\lambda^l$ is defined for each la    yer through "trust" coefficient $\eta < 1$:
\begin{equation}
   \lambda^l =  \eta  \times \frac{||w^l||}{||\nabla L(w^l)||}
    \label{eq:lars}
  \end{equation}
 The $\eta$ defines how much we trust the layer to change its weights during one update.
 
Note that now the magnitude of the update for each layer doesn't depend on the magnitude of the gradient anymore, so it helps to partially eliminate vanishing and exploding gradient problems. This definition  can be easily extended for SGD to balance the local learning rate and the weight decay term $\beta$:

\begin{equation}
    \lambda^l = \eta  \times \frac{||w^l||} {||\nabla L(w^l)|| + \beta *||w^    l|| }
   \label{eq:lars_wd}
\end{equation}

It has been observed that batch sizes can be scaled to a size of upto 64,000 whilst using LARS \cite{jia2018highly}. This is great for a system like Hydra which can have a large number of machines concurrently training a model, the more the number of machines, the larger the batch sizes possible. As Hydra needs to be latency sensitive, a mixed precision training pipeline is adopted. Mixed precision training uses half precision bits to train a network instead of using single precision bits. Using half precision bits increases the arithmetic throughput for some processors with optimized operations of lower bit data types. Mixed precision training has shown a decrease in memory consumption upto a factor of 8 \cite{micikevicius2017mixed}. It uses the following strategies during training:

\begin{itemize}
    \item Model weights and parameters are stored in a half precision format (16 bits) while computing the forward and backward pass. 
    \item That weight update rule is slightly modified. A master copy of the weights is kept in single precision format (32 bits). The gradients are converted to single precision format and weights are updated henceforth. This is because updating the weights in half precision leads to buffer overflows that result in 0 losses. After the new eights are computed, these weights are stored in memory and a half precision copy of these weights are used for the next forward and backpropagation iteration.
    \item While using half precision it is observed that the weights do not use half the representation capacity offered by the data type. This leads to gradient updates that are 0 which in turn the degree of convergence of the final model. A simple trick to solve this is to scale the loss \cite{lin2017deep}. Applying backpropagation after loss scaling scales the gradients into using the complete range of representational capacity of the data type.
\end{itemize}

Hydra also adopts gradient compression \cite{lin2017deep} using the DGC technique to compress gradients before communication to further boost performance on slow connections.

\section{Challenges and Future Work}

There are several challenges to building a system like Hydra due to the unreliable nature of it's nodes. Some of these challenges are listed below.
\begin{itemize}
\item If a majority of nodes drop off during some critical task such as training or tracker replication, there is a high probability of arriving at a deadlocked or inconsistent state where the task is stalled or data is lost.

\item The training time will be slower in comparison to cloud providers with dedicated data centers, also it might be less stable due to the unpredictable nature of the nodes.

\item Datasets can reduce in size if some peers are not online, since a majority of data will have only a single node replication. Better replication strategies can potentially evade this problem.

\item Data contributed to a dataset can be of uncertain validity inspite of validation checks. Duplicate data, irrelevant data, malicious data can be easily contributed through various hacks. As Hydra hinges it's system upon the validity of data and the amount of data contributed it is imperative to explore security policies that are more robust.

\item As Hydra has not been evaluated on data as yet, it is difficult to judge how efficient and accurate the system is. A set of experiments evaluating speed, stability and fault tolerance will give a clearer idea of Hydra's fallacies.

\item We can further look towards improving user incentivization by incorporating concepts of game theory, learning theory and social choice theory. Also a standard metric that correlates Hydra coin with user contributions needs to be further worked upon.

\item Federated learning offers several approaches to train on a cluster of low powered devices, one of them being the Federated Averaging Algorithm \cite{federated}. Incorporating elements of FL such as reduced data transfer by prioritising host peers of the dataset being used for training into the load distribution policy can be explored \cite{kahngincentivizing}.

\end{itemize}

Exploring policies for fault tolerance and security strategies through experiments are considered as future work. As one can see, Hydra offers a ripe field for further investigation on different elements of a decentralized and incentivized framework.

\section{Conclusion}

We present a novel system which decentralizes dataset collection and distributed training of neural networks. The features of this system promote the collection of diverse and validated data, hence creating a data ingestion pipeline for training high quality models. We also describe how a distributed training framework can be build using this decentralized network of unpredictable nodes taking into account the heterogeneous nature of devices, slow network conditions and unreliable nodes.



\end{document}